\documentclass[fleqn,usenatbib]{mnras}

\usepackage{newtxtext,newtxmath}

\usepackage[T1]{fontenc}

\DeclareRobustCommand{\VAN}[3]{#2}
\let\VANthebibliography\thebibliography
\def\thebibliography{\DeclareRobustCommand{\VAN}[3]{##3}\VANthebibliography}

\usepackage{graphicx}	
\usepackage{amsmath}	
\usepackage{bm}
\usepackage{xcolor}
\usepackage{soul}
\usepackage{tabularx}

\definecolor{orange}{rgb}{1,0.5,0}
\definecolor{sred}{rgb}{.5,0,0}

\newcommand{\dd}{\mathrm{d}}

\title[Approximate Atmospheric Scattering using PINNs]{Approximating Rayleigh Scattering in Exoplanetary Atmospheres using Physics-informed Neural Networks (PINNs)}

\author[D. Dahlb\"udding et al.]{
David Dahlb\"udding$^{1,2,3}$\thanks{E-Mail: ddahlb@mpe.mpg.de},
Karan Molaverdikhani$^{1,2}$,
Barbara Ercolano$^{1,2}$
and Tommaso Grassi$^{2,3}$
\\
$^{1}$Universitäts-Sternwarte, Fakultät für Physik, Ludwig-Maximilians-Universität München, Scheinerstr. 1, D-81679 München, Germany\\
$^{2}$Exzellenzcluster ‘Origins’, Boltzmannstr. 2, D-85748 Garching, Germany\\
$^{3}$Centre for Astrochemical Studies, Max-Planck-Institut für extraterrestrische Physik, Gießenbachstr. 1, D-85749 Garching, Germany
}

\date{Accepted 2024 July 30. Received 2024 July 8; in original form 2024 March 6}

\pubyear{2024}

\begin{document}
\label{firstpage}
\pagerange{\pageref{firstpage}--\pageref{lastpage}}
\maketitle

\begin{abstract}
This research introduces an innovative application of physics-informed neural networks (PINNs) to tackle the intricate challenges of radiative transfer (RT) modeling in exoplanetary atmospheres, with a special focus on efficiently handling scattering phenomena. Traditional RT models often simplify scattering as absorption, leading to inaccuracies. Our approach utilizes PINNs, noted for their ability to incorporate the governing differential equations of RT directly into their loss function, thus offering a more precise yet potentially fast modeling technique.
The core of our method involves the development of a parameterized PINN tailored for a modified RT equation, enhancing its adaptability to various atmospheric scenarios. We focus on RT in transiting exoplanet atmospheres using a simplified 1D isothermal model with pressure-dependent coefficients for absorption and Rayleigh scattering. In scenarios of pure absorption, the PINN demonstrates its effectiveness in predicting transmission spectra for diverse absorption profiles. For Rayleigh scattering, the network successfully computes the RT equation, addressing both direct and diffuse stellar light components.
While our preliminary results with simplified models are promising, indicating the potential of PINNs in improving RT calculations, we acknowledge the errors stemming from our approximations as well as the challenges in applying this technique to more complex atmospheric conditions. Specifically, extending our approach to atmospheres with intricate temperature-pressure profiles and varying scattering properties, such as those introduced by clouds and hazes, remains a significant area for future development.
\end{abstract}




\section{Introduction}

The use of Machine Learning (ML) in exoplanetary science has grown increasingly popular during past years \citep[][]{Nixon2020}. Hereby, it is mostly used to accelerate the time-consuming process of finding the posterior distributions of atmospheric parameters given some observed spectrum. Usually, this inference is done using a Bayesian scheme, like nested sampling \citep[e.g.][]{NestedSamplingOG} where many so-called forward models simulate data which can then be compared to observations.

Pioneered by \cite{Waldmann2016}, who used a simple multi-layer perceptron (MLP) to predict the existence of molecules in exoplanetary atmospheres from their transit spectra, the approaches have diversified since then, from Random Forests \citep[][]{MarquezNeila2018,Fisher2020,Nixon2020} to generative adversarial networks like ExoGAN \citep[][]{Zingales2018} and convolutional neural networks, often in combination with Monte Carlo dropout \citep[][]{Soboczenski2018,Cobb2019,ArdevolMartinez2022}.

More recent approaches achieve generating more accurate and versatile posterior distributions by employing normalizing flows in combination with variational inference \citep[][]{Yip2022}, neural posterior estimation \citep[][]{Vasist2023}, or flow matching and importance sampling \citep[][]{Gebhard2023}.

The approaches most similar to our own are by \cite{Himes2022} and \cite{Hendrix2023}, who both try to speed up the forward model while leaving the inference scheme untouched. While \cite{Himes2022} use a neural network to emulate radiative transfer by predicting the spectrum given some atmospheric parameters, \cite{Hendrix2023} focus on accelerating computationally expensive disequilibrium chemistry calculations using an autoencoder and an LSTM-like neural network.

Most of these models rely on traditional numerical methods to generate the data that the ML model uses for training. Since all ML models cannot contain more information than their training data, they cannot generalize well beyond the used forward model.
Including the time to train a model, the respective method only offers a speed-up if enough exoplanets can be analyzed using the same model trained on the same data. This can be particularly challenging for exoplanetary atmospheres, given the diversity of exoplanets that have been observed. Additionally, as soon as a forward model is updated to include some new insight, the ML model might be completely retrained. It can also prove difficult to use one model for observations with varying instruments, that, e.g., have a different number of data points or an incompatible spectral range.

In the future, as data become more abundant and precise and the models more detailed, the computational burden of running a retrieval will further increase. Therefore, in this work, we will assess the feasibility of using pre-trained physics-informed neural networks (PINNs) to cheaply model complex phenomena in exoplanetary atmospheres. We focus on the simple example of Rayleigh scattering in the isothermal atmosphere of a transiting exoplanet.

Building upon the work of \cite{Mishra2021}, who used PINNs to solve the radiative transfer (RT) equation, we first model RT considering only absorption. Since for transiting exoplanets, the traditional numerical method treats scattering the same as absorption in that the light scattered out of the beam is analogous to absorbed light \citep[e.g.][]{pRT}, we can compare our PINN to existing models to establish its speed and accuracy.

Secondly, we include Rayleigh scattering in the RT equation. As has been shown by \cite{Sengupta2020}, scattered light can have a significant effect on the transit spectrum at short wavelengths. Using PINNs could help us study this effect in more detail and potentially improve our estimates of the hydrogen-to-helium ratio or the existence and properties of clouds and hazes.

In Section 2, after briefly introducing PINNs, we discuss how the respective problem is presented to both the absorption and scattering PINN, such that they are able to solve it efficiently. This involves generating and normalizing the inputs, the formulation of the loss function, and the exact training procedure.
The results of both models are assessed in Section 3, while drawbacks are summarized in Section 4. Section 5 discusses key insights and possible enhancements.

\section{Methods}

\subsection{PINNs}

While there are many different ways of incorporating physical laws into the training of a neural network, PINNs, as introduced by \cite{Raissi2019PINN}, achieve this through their loss function, which is directly based on a differential equation.

To see how a PINN is able to solve a nonlinear partial differential equation (PDE), as explained in \cite{Cuomo2022ScientificML}, let us first describe a PDE in its most general form defined on the domain $\Omega \in \mathbb{R}^\mathrm{d}$ with its boundary $\partial \Omega$:

\begin{align}
    \mathcal{F}(u(z);\gamma) = f(z), \quad z \in \Omega \label{eq:diffEq} \\
    \mathcal{B}(u(z)) = g(z), \quad z \in \partial \label{eq:boundaryCond} \Omega\,,
\end{align}
\\
where $u$ describes the unknown solution to the differential equation, dependent on the space-time coordinate vector $z := [x_1, \cdots, x_\mathrm{d-1}, t]$. $\mathcal{F}$ refers to the nonlinear differential operator dependent on physics-related parameters $\gamma$, while $f$ is the function identifying the data of the problem (if available). Eq. (\ref{eq:boundaryCond}) defines the boundary and initial conditions, where $\mathcal{B}$ is again an operator and $g$ is the boundary function.

While the PINN can also be used for an inverse problem of finding the parameters $\gamma$, we use it for approximating the solution $u$ for every $z$, dependent on the (given) parameters $\gamma$. Hereby, the PINN itself functions as the approximation to this solution. The PINN therefore has inputs $z$ and output $\hat u_\theta(z) \approx u(z)$ dependent on the network parameters $\theta$. The corresponding loss function that should be minimized is made up of different components $\mathcal{L}_i$ with respective weights $\omega_i$:

\begin{equation}
    \mathcal{L} = \omega_\mathcal{F} \mathcal{L}_\mathcal{F} + \omega_\mathcal{B} \mathcal{L}_\mathcal{B}\,.
    \label{eq:lossComponents}
\end{equation}
\\
The first term $\mathcal{L}$ is the differential equation (\ref{eq:diffEq}) itself, with all terms on one side of the equation such that it is equal to zero. Often, the following form analogous to the mean squared error is used:

\begin{equation}
    \mathcal{L}_\mathcal{F} = \sum_{i=1}^n \left( \mathcal{F} (\hat u_\theta(z_i);\gamma) - f(z_i) \right)^2, \quad z_i \in \Omega\,,
    \label{eq:residualLoss}
\end{equation}
\\
where the square guarantees both positive values, such that the minimum lies at zero, and differentiability (compared to the mean absolute error).

Now we calculate the value of the differential equation $\mathcal{F}(u(z);\gamma)$ with automatic differentiation. We can not only use it to calculate the gradient of the loss function with respect to the weights of the network, but also calculate the gradient of the output $\hat u_\theta(z)$ of the network with respect to the inputs $z$, the space and time coordinates. Hence, we can insert all the values of the variables and derivatives into the differential equation to calculate the loss.

The second term can be calculated accordingly to make sure that the boundary conditions are fulfilled:

\begin{equation}
    \mathcal{L}_\mathcal{B} = \sum_{i=1}^n \left( \mathcal{B} (\hat u_\theta(z_i)) - g(z_i) \right)^2, \quad z_i \in \partial \Omega\,.
    \label{eq:boundaryLoss}
\end{equation}
\\
In this work, our approach to employing PINNs deviates from their conventional application. Typically, PINNs are tasked with solving specific instances of differential equations, where their advantage over traditional numerical methods is most pronounced in scenarios involving complex geometries or boundary conditions. However, we leverage PINNs in a novel way by training them on a parameterized form of the radiative transfer equation. This means that the solution approximation provided by the PINN, denoted as $\hat u_\theta(z)$, is influenced not only by the space-time coordinates $z$ but also by a set of physical parameters $\gamma$. This unique application allows the parameterized PINN to generate solutions, $\hat u_\theta(z, \gamma)$, for varying physical conditions without the need for retraining, showcasing a flexible and powerful tool for modeling radiative transfer processes in exoplanetary atmospheres.

\begin{figure*}
\begin{center}
   \begin{minipage}[h]{.65\textwidth} 
      \includegraphics[width=\linewidth]{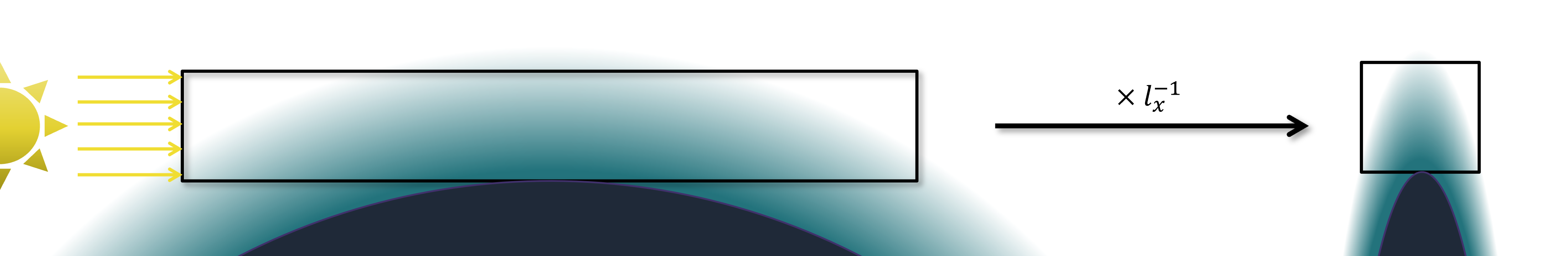}
   \end{minipage}
   \hspace{.05\linewidth}
   \begin{minipage}[h]{.25\linewidth} 
      \includegraphics[width=\linewidth]{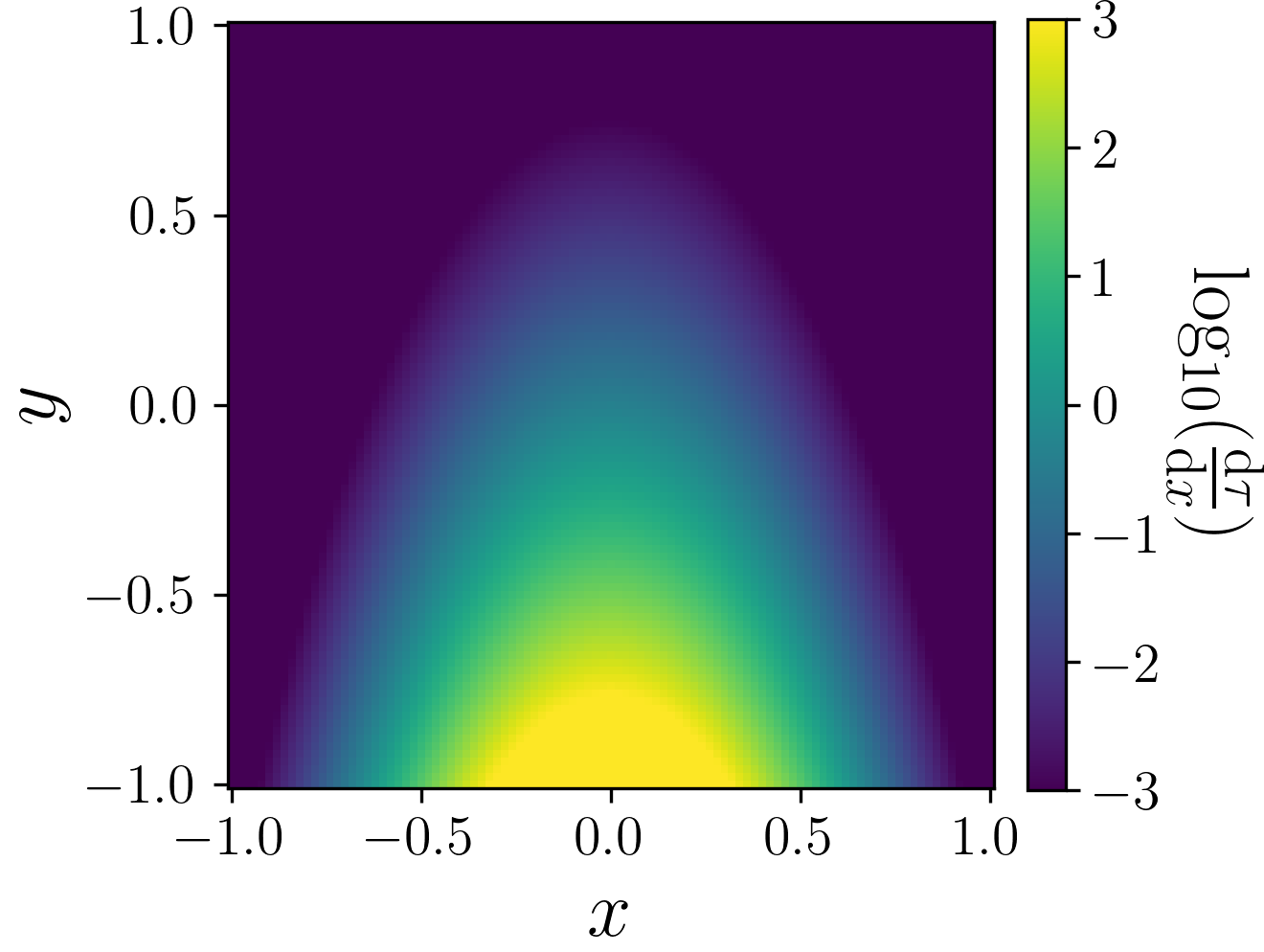}
   \end{minipage}
   \caption{Left: The transit geometry. The x (and y) axis was scaled by a factor of $l_x$ (and $l_y$), such that both coordinates are normalized.\\
   Right: Example differential optical depth $\frac{\dd \tau}{\dd x} = l_x \alpha(x,y)$, where $\alpha \sim P$.}
   \label{fig:geometry}
\end{center}
\end{figure*}

\subsection{Setup} \label{sec:setup}

In this study, our objective extends beyond solving a PDE; we aim to tackle an integro-differential equation, specifically the stationary radiative transfer equation, which operates in two spatial dimensions:

\begin{equation}
\begin{split}
    \mathrm{cos}(\phi) \frac{\dd u}{\dd x} &+ \mathrm{sin}(\phi) \frac{\dd u}{\dd y} + (\alpha_a + \alpha_s) \cdot u\\
    - \frac{\alpha_s}{4\pi} &\cdot \int \Phi(\theta, \phi, \theta', \phi') u(\theta', \phi') \dd\Omega' = 0\,,
    \label{eq:rte}
\end{split}
\end{equation}
\\
where the angle $\phi$ = 0 is defined as the positive x-direction and lies
in the range $\phi \in [-\pi, \pi]$. $\alpha_a$ and $\alpha_s$ are the absorption and scattering coefficient, respectively and $\Phi$ the scattering phase function, depending on the incoming ($\theta', \phi'$) and outgoing angles ($\theta, \phi$) of the light. In the case of Rayleigh scattering, the scattering phase function depends only on the angular distance $\Delta$:

\begin{equation}
    \Phi(\Delta) = \frac{3}{4} \left( 1 + \mathrm{cos}^2\Delta \right)\,,
    \label{eq:scatteringPhaseFunction}
\end{equation}
\\
where $\mathrm{cos}\Delta = \mathrm{cos}\theta \, \mathrm{cos}\theta' + \mathrm{sin}\theta \, \mathrm{sin}\theta' \, \mathrm{cos}(\phi - \phi')$ according to the law of cosines. For both the absorbed and scattered light component sensible symmetries for small angular extents of the star are exploited such that the integral can be evaluated numerically over only one angle (see Appendix \ref{app:sca_int} for the explicit integral).

The function of interest that we want to solve for is $u(x,y,\phi)$, the transmission through the atmosphere. For this, we also need to define our geometry, as shown in Fig.\,\ref{fig:geometry}. More specifically, we want to find the transmission $u(x=x_\mathrm{max},y,\phi=0)$, which we can then integrate over the $y$ or radial direction to find the effective area of the planet.

This depends on a number of other parameters. So, to reduce the dimension of the problem, we assume an isothermal atmosphere and constant conditions throughout the day-night-terminator region of the planet. The models are trained on an atmosphere spanning pressures from $10^{-6}$ to $10^2$ bar and consisting of 100 log-spaced layers, where the absorption and scattering coefficient is interpolated logarithmically between them.\\
Furthermore, for the PINN with scattering, we assume that the planet transits its star with an orbital inclination $i = 90^\circ$, simplifying the boundary condition. The scattering PINN also only calculates the solution for pure scattering without any absorption ($\alpha_a = 0$). The total transmission function, when simulating a spectrum, is then calculated by multiplying the solution for purely absorbing species with the solution for purely scattering species, which should hold as long as one of the two is dominating.

\subsection{Data Generation} \label{sec:dataGen}

Before we are able to train a neural network on this problem, we have to generate and normalize the necessary input data.\\
For both the absorption and the scattering model, this includes the spacial coordinates $x$ and $y$, which are normalized to be within the interval $[-1,1]$ by scaling them with factors $l_x$ and $l_y$, respectively. These are defined such that the innermost atmospheric layer touches the lower boundary at $(x,y) = (0, -1)$ and the outermost layer touches the upper boundary at $(x,y) = (0, 1)$ as well as going through the lower corners $(x,y) = (-1, -1)$ and $(x,y) = (1, -1)$:

\begin{equation}
    l_x = \sqrt{r_0^2 - r_{99}^2}, \quad l_y = \frac{r_0 - r_{99}}{2}\,,
\end{equation}

where $r_0$ refers to the outermost and $r_{99}$ to the innermost layer. Note that we have to account for these scaling factors inside of the differential equation by dividing the respective gradients by this factor.

The other two input parameters which are given to both models include the radius of the planet $R_0$ at the reference pressure $P_0 = 10^{-2}$ bar, spanning from $0.2 R_\mathrm{J}$ to $2.0 R_\mathrm{J}$, normalized to also be within $[-1,1]$. Secondly, the scale height $H := k_{\rm B} T / (\mu \, g)$, where T refers to the temperature, $\mu$ to the mean molecular weight, and g is the gravity at the reference pressure $P_0$. $\log_{10}(H[{\rm cm}])$ is sampled from a normal distribution with a mean of 7.1 and a standard deviation of 0.35 and is normalized accordingly. Extreme values of the scale height can change the geometry of the problem significantly and hence, make it harder for the PINN to learn a solution. The normal distribution ensures that this does not hinder the training of the PINN, while maintaining a broad range of values in the training data, allowing the neural network to approximate regions of higher data density more quickly and accurately \citep[e.g.,][]{Basri2020FrequencyBias}. Together, $H$ and $R_0$ allow us to calculate the radius profile $r(P)$ assuming a constant temperature but variable gravity:

\begin{equation}
    r(P) = R_0 \cdot \left[ 1 + \frac{H}{R_0} \mathrm{ln} \left(\frac{P}{P_0}\right) \right]^{-1}\,.
\end{equation}

Although isothermality is often an inadequate approximation for exoplanetary atmospheres \citep[e.g.,][]{Rocchetto2016}, we use it to assess if the PINN can manage the simplest scenario. That is why we further assume constant mass fractions (or mixing ratios) of all species throughout the atmosphere.

For the pure-absorption PINN, we also need to generate the $\alpha_a(P)$-profiles, where the absorption coefficient $\alpha_a$ is extended to encompass both absorption and scattering sources. In this approach, scattering is effectively treated as a form of absorption, consistent with the traditional model. Ideally, these profiles should not just consist of a database of absorption coefficients from certain molecules at certain wavelengths, but should be more general by generating them randomly on the fly during training. For this, we first investigated example $\alpha_a(P)$-profiles using petitRADTRANS \citep[][]{pRT}, which calculates the absorption coefficient $\alpha_a = \kappa X \rho$ for every species at every layer as the product of opacity $\kappa$, mass fraction $X$ and density $\rho$. Our analysis included the correlated-k opacities of H$_2$O and CO \citep[][]{Rothman2010HITEMP}, CH$_4$ and CO$_2$ \citep[][]{Chubb2021ExoMol} and Na and K \citep[][]{Piskunov1995VALD}, as well as Rayleigh scattering from H$_2$ \citep[][]{Dalgarno1962H2Rayleigh} and He \citep[][]{Chan1965HeRayleigh} and collisionally-induced absorption from H$_2$-H$_2$ \citep[][]{Borysow2001H2H2_1, Borysow2002H2H2_2} and H$_2$-He \citep[][]{Borysow1988H2He_1, Borysow1989H2He_2, Borysow1989H2He_3}. We found that most profiles follow either $\alpha_a$ $\sim \rho \sim P$ or $\alpha_a \sim P^2$ (if $\kappa \sim P$) and hence $\frac{\dd\ln(\alpha_a)}{\dd\ln(P)}$ either equates to 1 or 2. Therefore, a truncated exponential profile around 1 and 2 ($\leq 2$) was used to create the profiles from their logarithmic gradient. During training, we observed that the PINN improves when using more accurate but diverse profiles. Therefore, we include 6 different types of randomly generated profiles in total:

\begin{itemize}
    \item $\frac{\dd\ln(\alpha_a)}{\dd\ln(P)} \approx 1$
    \item $\frac{\dd\ln(\alpha_a)}{\dd\ln(P)} \approx 2$
    \item random transition from $\frac{\dd\ln(\alpha_a)}{\dd\ln(P)} \approx 1$ to 2
    \item random transition from $\frac{\dd\ln(\alpha_a)}{\dd\ln(P)} \approx 2$ to 1
    \item randomly switching between $\frac{\dd\ln(\alpha_a)}{\dd\ln(P)} \approx 1$ and $\frac{\dd\ln(\alpha_a)}{\dd\ln(P)} \approx 2$
    \item random negative values ($\geq -0.5$) for $\frac{\dd\ln(\alpha_a)}{\dd\ln(P)}$ in some profiles.
\end{itemize}

\begin{figure}
    \centering
    \includegraphics[width=\columnwidth]{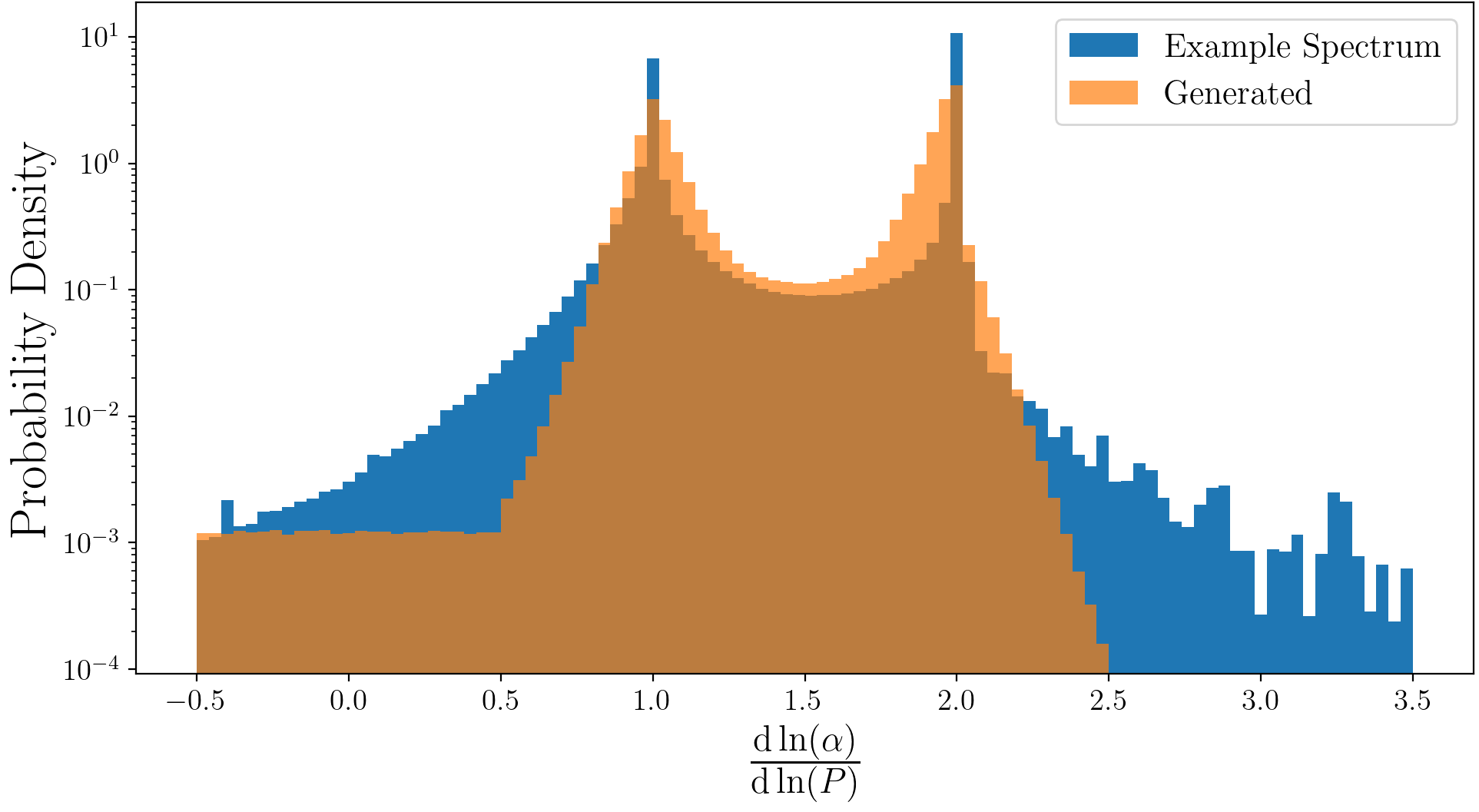}
    \caption{Testing the $\alpha_a(P)$-profile generation algorithm by comparing the resulting distribution of logarithmic gradients between every layer with the same distribution for all opacity sources of the example spectrum shown in Fig. \ref{fig:exampleSpec_absPINN}. The portion of gradients of the example spectrum that lay outside of the shown interval accumulate to less than 0.1\% for $\frac{\dd\ln(\alpha_a)}{\dd\ln(P)} < -0.5$ and to about 1.1\% for $\frac{\dd\ln(\alpha_a)}{\dd\ln(P)} > 3.5$.}
    \label{fig:alpha-grad-hist}
\end{figure}

In Fig. \ref{fig:alpha-grad-hist}, we compare the distribution of the generated logarithmic gradients with actual ones from all opacity sources and layers of the example atmosphere shown in Fig. \ref{fig:exampleSpec_absPINN}. The more diverse set of gradient-profiles improved the error of a spectrum from initially around 3-5\% to 1-2\%.

The final $\alpha_a(P)$-profiles are then created by cumulatively summing up the gradients in both directions from a middle layer (with index 49), which gets assigned a $\log_{10}(\alpha_a)$-value by a uniform distribution and hence acts as a random shift of the profiles.

To inform the PINN about the profile, we give it all the absorption coefficients $\alpha_i, i \in [0,99]$ from the 100 atmospheric layers, such that it has a total of 104 inputs.

In the case of Rayleigh scattering from hydrogen and helium, the scattering cross-section does not depend on the pressure. Hence, the scattering coefficient is simply proportional to the density $\alpha \sim \rho \sim P$, an example of which can be seen in Fig. \ref{fig:geometry}, and can be parameterized with just one input variable.

The scattering PINN has additional inputs, namely the angle $\phi \in [-\pi, \pi]$ as well as (half) the angular extent of the star as seen from the planet $\Delta_* \in [0, 0.3\pi]$. Both are normalized to be within $[-1,1]$.
A summary of the parameter bounds and how they are sampled can be found in Table \ref{tab:input_ranges}.

\begin{table}
    \caption{Ranges of the input parameters to the absorption and scattering PINN. Most are sampled uniformly within their respective intervals, with the exceptions of the angle $\phi$, half of which is sampled uniformly within $[-\Delta_*,\Delta_*]$, the scale height $H$, which is sampled via a normal distribution in log-space, and the absorption profile $\alpha_a$ (see the Section \ref{sec:dataGen} for details on the generation of the $\alpha_a(P)$-profiles). The 49 refers to the absorption or scattering coefficient at the layer with index 49 (ranging from 0 to 99).}
    \label{tab:input_ranges}
    \begin{tabularx}{\columnwidth}{XX}
    \hline
    Input & Range \\
    \hline
    \hline
    $x$ & $[-l_x, l_x]$ \\
    $y$ & $[-l_y, l_y]$ \\
    $\phi^\dag$ & $[-\pi, \pi]$ \\
    $R_\mathrm{P} [R_\mathrm{J}]$ & $[0.2,2.0]$ \\
    $\log_{10}(H[\mathrm{cm}])$ & $\mathcal{N}(7.1, 0.35^2)$ \\
    $\log_{10}(\alpha_{a,49} [\mathrm{cm}^{-1}])^\ast$ & $[-20.2,-4.5]$ \\
    $\log_{10}(\alpha_{s,49} [\mathrm{cm}^{-1}])^\dag$ & $[-18.74,-6.74]$ \\
    $\Delta_*^\dag$ & $[0,0.3\pi]$ \\
    \hline
    \multicolumn{2}{l}{$^\ast$: input only to absorption PINN}\\
    \multicolumn{2}{l}{$^\dag$: input only to scattering PINN}
    \end{tabularx}
\end{table}

\subsection{Training}

The construction of the loss function for the absorption PINN initially appears straightforward, but involves intricate considerations. The boundary condition is established by the incident stellar radiation, characterized by the function \( u(x=-1, y) = 1.0 \), representing the intensity of starlight entering the system from the left boundary. This boundary condition is coupled with the RT equation governing the system:

\begin{equation}
    \frac{\dd u}{\dd x} + \alpha' \cdot u = 0\,,
    \label{eq:absRTE}
\end{equation}

where the scaling factor is taken into account inside the $\alpha' \coloneqq l_x \alpha_a$.
However, utilizing this equation directly as the loss function presents significant challenges in solving the problem. To understand why this is the case, let us consider two extremes for different regions of the atmosphere: $\alpha' \ll 1$ and $\alpha' \gg 1$.

For an extremely high value of $\alpha'$, the RT equation effectively becomes $\alpha' \cdot u = 0$ with the simple solution $u = 0$. Now consider small deviations of $u$ from 0; Because of the high $\alpha'$ they can result in high values of the loss function and proportionally steep gradients. This means that the PINN effectively overfits these regions to be as close to zero as possible, even though other regions might be of more interest. We can solve this by dividing the whole equation by $\alpha'$. This gives us approximately $u = 0$ as our loss function in such regions, resulting in the PINN learning the right solution without overfitting.

In the second case where $\alpha' \ll 1$, the form of Eq.\,(\ref{eq:absRTE}) would be preferable, since it would effectively be $\dd u/\dd x = 0$. Again, the neural network would learn the right solution of $u = \mathrm{const.}$, but would not overfit the region as would be the case if we would divide the gradient by $\alpha'$.

We can combine these two forms of the same equation by calculating both and taking the (element-wise) minimum.
The RT equation, which only considers Rayleigh scattering, is normalized in a similar fashion, details of which can be found in Appendix \ref{app:res_loss}.

Meanwhile, the boundary condition of the scattering PINN has more complexity to it. On the left side, the light from the central star enters the atmosphere, this time with an angular extent. Because of this angular extent, also light from the upper boundary could enter the atmosphere, while from the right boundary, no light should enter but only leave the considered slab:

\begin{equation}
\begin{split}
    u(x=-1, y, |\phi|\leq\frac{\pi}{2}) &= 1 \; \mathrm{if} \; |\phi|\leq\Delta_*, \mathrm{else} =0\\
    u(x, y=1, \phi \leq 0 ) &= 1 \; \mathrm{if} \; |\phi|\leq\Delta_*, \mathrm{else} =0\\
    u(x=1, y, |\phi|\geq\frac{\pi}{2}) &= 0\,.
\end{split}
\end{equation}

\begin{figure}
    \centering
    \includegraphics[width=\columnwidth]{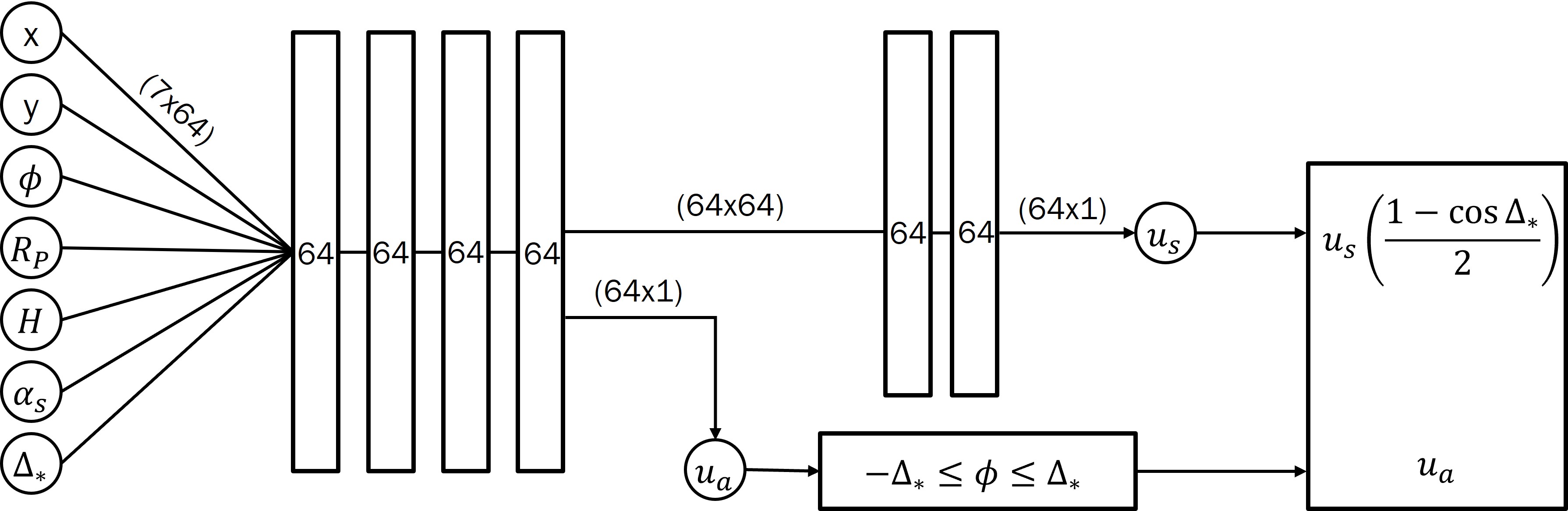}
    \caption{The architecture of the scattering PINN, where the numbers in the (fully-connected) layers indicate their width. The 4th layer connects both to an output $u_a$ and to two subsequent layers that have an output $u_s$. Because $u_a$ models the part of the solution that fulfills the RTE without scattering, a filter is applied, dependent on $\phi$, such that $u_a = 0$ where the light cannot get without changing directions. The combined output $u_a + u_s$ should fulfill the complete RTE, such that $u_s$ only has to model the scattered light component of the solution. Since the diffuse light only becomes relevant for large enough $\Delta_*$, $u_s$ is multiplied with the solid angle subtended by the star.}
    \label{fig:ScaPINN-architecture}
\end{figure}

For the lower boundary, defining an exact boundary condition poses a challenge due to the uncertainty in the amount of light that could be scattered into the observed atmospheric slab from below. However, the absence of a boundary condition can potentially lead to erroneous solutions. To mitigate this issue, we approximate the lower boundary condition by setting the scattered light component to zero. This implies that no additional scattered light enters the system from this boundary. This approximation is expected to yield an accurate solution, provided that the optical depth of the atmosphere is sufficiently high.

For the absorption component, an accurate boundary condition can be found by recognizing the spherical symmetry of the system and rotating the cord of the light ray along a path of constant radius until the point of interest reaches the x-coordinate of $x=0$:

\begin{equation}
\begin{split}
    &u_a(x_0 \leq 0,y=-1,\phi \geq 0) = u_a(x=0,y',\phi+\Delta\phi),\\ 
    &\mathrm{where} \; y' = y\left(x_0, r(x_0,y=-1)\right), \; \mathrm{tan}(\Delta\phi) = \frac{l_x x_0}{r_{99}}.
\end{split}
\end{equation}

Note that this only works for $x \leq 0$. Otherwise, this boundary condition is set to zero, which again should hold, if the optical depth of the atmosphere is high enough at the highest pressures. The other edge case that we have to consider is if $\mathrm{abs}(\phi+\Delta\phi) > \Delta_{*,\mathrm{max}} (= 0.3 \pi)$, for which $u_a$ is zero because of the inherent architecture of the PINN. As this only happens for particularly low values of $r_{99}$ and large values of $|l_x x_0|$ (i.e., for the outer layers of the atmosphere), the boundary condition is set to $u_a = 1$ for these rare instances.

The final boundary condition is to ensure the periodicity of $\phi$: $u(x,y,\phi=\pi) = u(x,y,\phi=-\pi)$.

Because of its boundary conditions, the scattering PINN has a special architecture, as shown in Fig.\,\ref{fig:ScaPINN-architecture}. This architecture has two outputs: One ($u_a$) should fulfill the RTE without scattering and utilizes a hard-coded filter, such that the neural network does not have to learn the non-differentiable boundary at $|\phi| = \Delta_*$, where the value of $u$ can jump from 1 to 0. The other output ($u_s$) should then model the diffuse component of the light such that when adding the two outputs $u = (u_a + u_s)$, they together fulfill the complete RTE with scattering.

If we would now sample $\phi$ uniformly in $[-\pi,\pi]$, most of the "absorption output" $u_a$ would equate to zero only because of the mentioned filter. This not only wastes a lot of training points, but also effectively lowers the weight (or importance) of the corresponding residual in the loss term. We mitigate this problem by sampling half of $\phi$ within $[-\Delta_*, \Delta_*]$.

Both the boundary conditions and the residuals enter the loss function with their mean square. To make sure that the boundary conditions are fulfilled first, a weight of 0.5 was given to the residual term ($\omega_\mathcal{F} = 0.5$ and $\omega_\mathcal{B} = 1$ in Eq.\,(\ref{eq:lossComponents})). The lower boundary condition is given a weight of 0.1, as it is reasonable to fulfill only if $u_a$ at $x=0$ is already sufficiently accurate. The final loss function is then the logarithm of their weighted sum.

The used optimizer is the higher-order quasi-Newton L-BFGS. We found that initially using Adam and then switching to L-BFGS yielded no benefit, so L-BFGS was used from the start. The parameters of the optimizer differ from the default \textsc{PyTorch} parameters in a learning rate of 0.8, the maximal number of both function evaluations and iterations per optimization steps of 50 (or 20), the tolerance change is set to the machine epsilon and the Strong Wolfe condition is chosen for the line search. As described in many PINN papers before \citep[e.g.][]{Sankaran2022}, we confirmed that having a larger batch size generally improves the final model. Accordingly, we always used the maximum batch size possible on an NVIDIA A100-SXM4-40GB GPU, with double the number of points for the residual loss compared to the boundary loss. To mitigate overfitting, the batch was resampled every 50 training steps (every 20 steps for the first 8000 steps in the case of the scattering PINN). The parameters that are computationally more expensive to sample (mainly the $\alpha_a$-profiles, but also the two radius profile parameters) were sampled pre-training with a size of $10^7$. During training, these parameters were then sampled from this saved array and loaded onto the GPU, thereby accelerating the process.

Throughout the training of the absorption PINNs, the inputs were resampled 100 times. Considering the number of points for which the residual loss is computed ($2^{21}$ or $2^{22}$), the total number of input combinations amounts to roughly $2.1 \times 10^8$ or $4.2 \times 10^8$, depending on the network size. For the scattering PINN, which was resampled 400 times and repeated thereafter, with $2^{19}$ residual points, the model was similarly trained on approximately $2.1 \times 10^8$ unique input combinations.

For the absorption PINN, a hyperparameter grid search is used to optimize the number of layers (4 or 6) as well as the width of each layer (32, 64, or 128). A PINN with more free parameters may improve our result even more, but the 6x128 network already needed more time to calculate a spectrum than the classical numerical method, so we decided against trying out a deeper or wider network. For each configuration, 7 PINNs were trained with different random Xavier initialization. They were evaluated on 95 test spectra with the help of petitRADTRANS \citep[][]{pRT}.

The number of subsequent layers of the scattering PINN were not optimized, although one layer was insufficient to accurately model the correct solution.

\section{Results}

\subsection{Absorption PINN}

\begin{figure}
    \centering
    \includegraphics[width=\columnwidth]{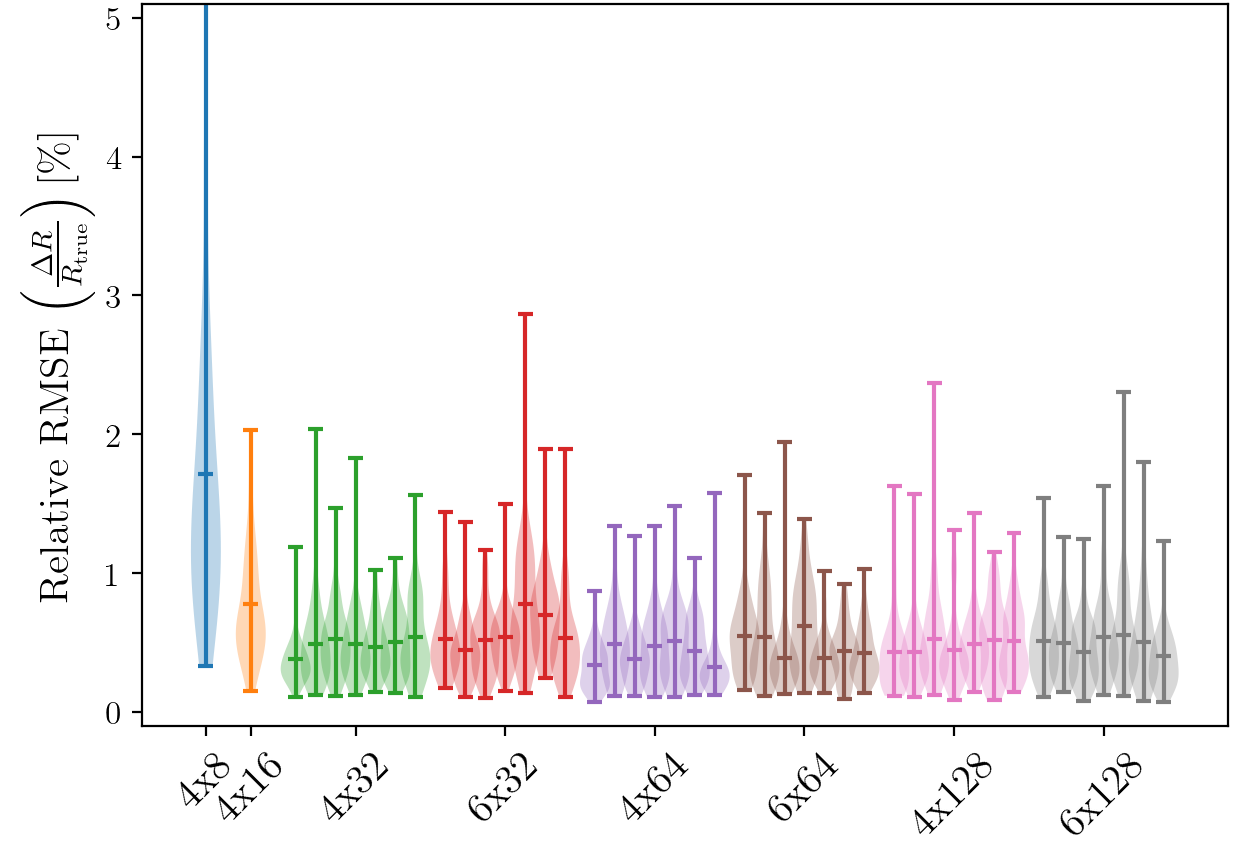}
    \caption{Violin plot of the 95 root mean squared errors of each test spectrum for each trained absorption PINN. The maximum RMSE of the 4x8 PINN lies outside the shown range at 7.7\%.}
    \label{fig:hp-search-errors}
\end{figure}

\begin{figure}
    \centering
    \includegraphics[width=\columnwidth]{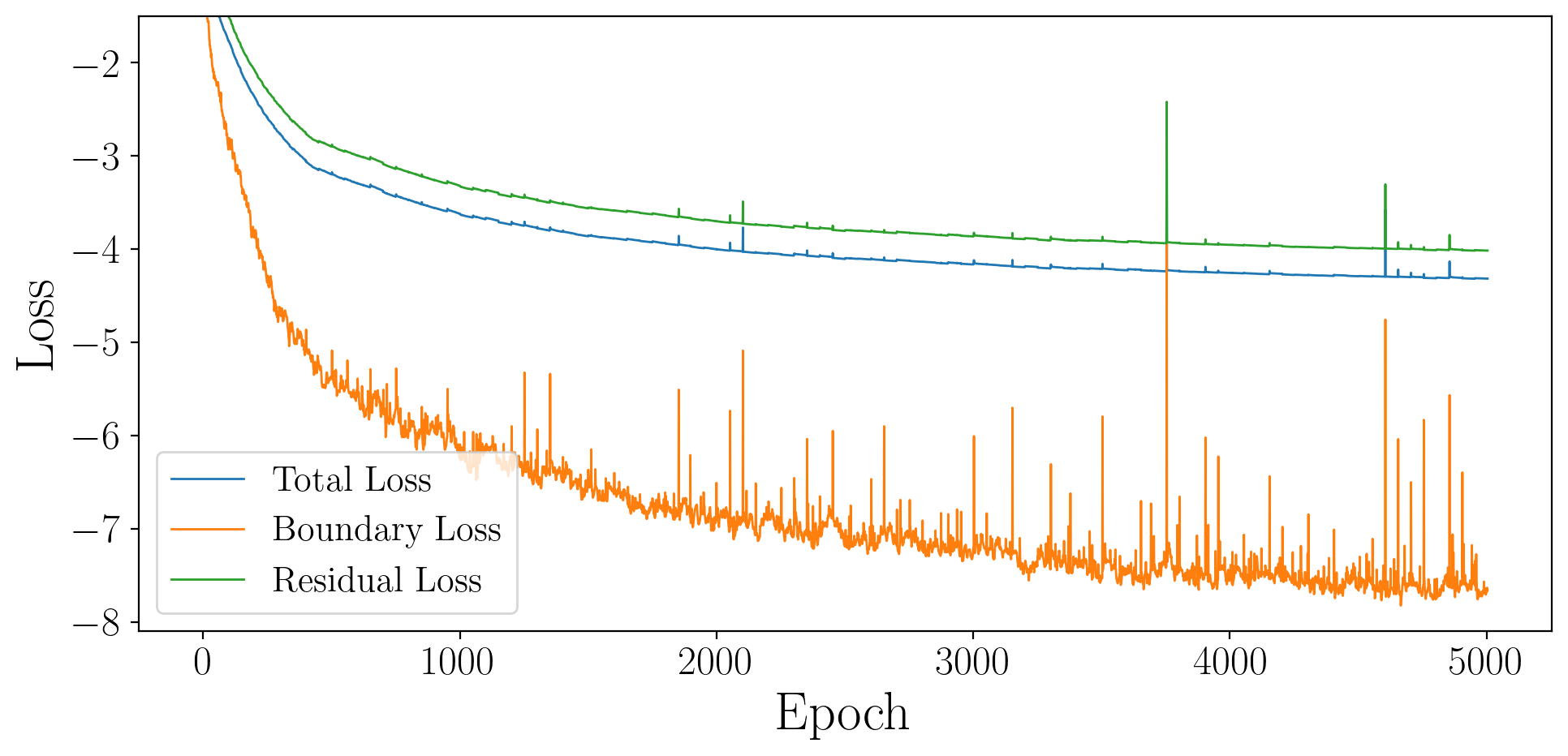}
    \caption{Example for the evolution of the different loss components, for the 3rd 4x64 PINN. The total loss $\mathcal{L}$ is calculated from the boundary ($\mathcal{L}_\mathcal{B}$) and residual loss ($\mathcal{L}_\mathcal{F}$) as $\log_{10}(\mathcal{L}) = \log_{10}(\mathcal{L}_\mathcal{B} + 0.5 \mathcal{L}_\mathcal{F})$, as defined in Eq.\,\ref{eq:lossComponents} - \ref{eq:boundaryLoss}.}
    \label{fig:lossComponents}
\end{figure}

\begin{figure}
    \centering
    \includegraphics[width=\columnwidth]{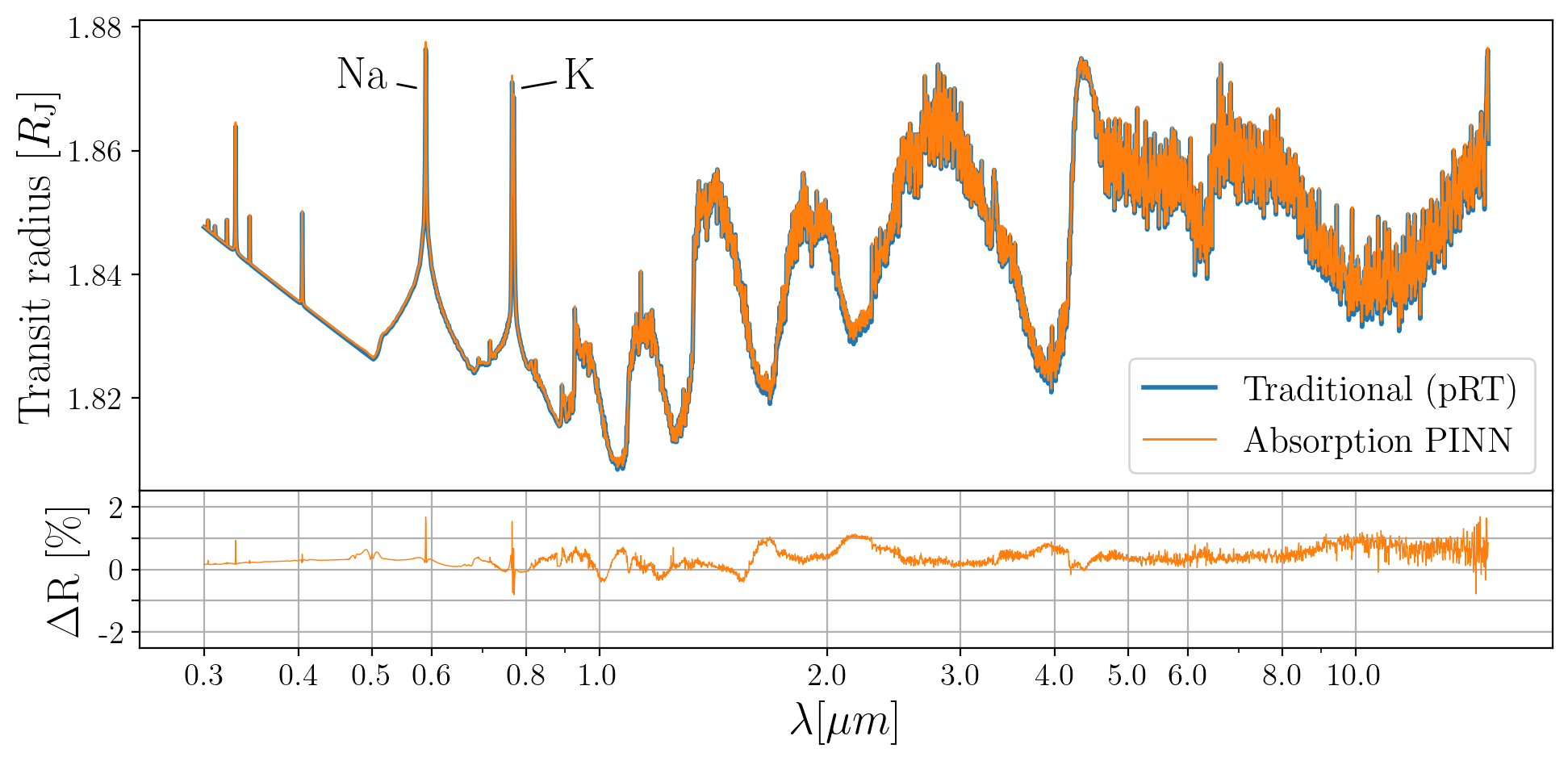}
    \caption{Example spectrum for one of the best-performing absorption PINNs, the 3rd 4x64 PINN. A hydrogen- and helium-dominated atmosphere, containing H$_2$O, CO, CO$_2$, CH$_4$, Na, and K.}
    \label{fig:exampleSpec_absPINN}
\end{figure}

After trying out what maximum batch size is possible for each hyperparameter configuration, we found that the three models with the most adjustable parameters (6x64, 4x128, 6x128) are able to utilize a batch size of $3 \times 2^{20}$ while the three smaller models (4x32, 6x32, 4x64) could use double the batch size ($3 \times 2^{21}$).

The distribution of the root mean squared errors with respect to each example spectrum can be seen in Fig.\,\ref{fig:hp-search-errors}. It seems that the 4x64 model consistently gives us the best result, although no architecture performs significantly worse.

Given that the errors of the smallest networks with a width of 32 were not significantly higher than those of larger networks, we experimented with an even smaller architecture of 4x16 and 4x8. For the 4x16 PINN, this adjustment resulted in a slight increase in error, similar to the worst-performing 6x32 network, and a considerably larger error in the case of the 4x8 PINN, but it produced diminishing returns in terms of speed.

For one of the 4x64 models, we also show the evolution of the different loss components during training in Fig\,\ref{fig:lossComponents}. Here we see how the weight $\omega_\mathcal{F}$ of 0.5 allows the network to quickly fulfill the boundary condition, while still giving the residual loss component sufficient weight to decrease over the epochs. The final value of this residual loss $\log_{10}(\mathcal{L}_\mathcal{F}) = -4.0$ already gives us a first order of estimate for the errors of the PINN of $10^{-2}$.

We also checked for possible correlations of the error with any atmospheric parameters. We only found that for really low radii, our formula for the radius profile $r(P)$ breaks down for the defined pressure range, yielding inaccurate results.

Fig.\,\ref{fig:exampleSpec_absPINN} presents an example spectrum, notably incorporating sodium (Na) and potassium (K). These elements serve as robust indicators for assessing the PINN's capacity to generalize beyond its initial training data, due to their potential for exhibiting abrupt increases in absorption coefficients at specific atmospheric layers. During the training phase, the $\alpha_a(P)$ profiles we generated typically did not showcase logarithmic gradients $\frac{\dd\ln(\alpha_a)}{\dd\ln(P)} > 2$. However, for these atomic species, the gradient between two layers can surge to values well above 2, occasionally surpassing 10. Such dramatic shifts can result in the optical depths of subsequent layers abruptly increasing from negligible to substantial values. Although $\alpha_a(P)$ profiles with such non-differentiable characteristics were not included in the training due to the complexity they introduced in minimizing the residual loss, the PINN demonstrates a commendable ability to approximate accurate solutions, even though this results in slightly larger errors than for known $\alpha_a(P)$ profiles. This success underscores the network's adaptability and effectiveness, showcasing its capability to handle complex scenarios that extend beyond the initial training data.

While the observed accuracy gives us confidence in our method, it does not offer any improvement in performance. Using the larger models of width 128 can even significantly slow down the simulation of a transit spectrum compared to the numerical method.

This contrasts with the results of \cite{Himes2022}, who could significantly speed up their forward model using a neural network. But their method directly calculates the spectrum (in a fixed spectral range and spectral resolution) from the atmospheric parameters (also fixed in their number and range). Our PINN, on the other hand, calculates the transmission function $u(y)$ for every individual $\alpha_a(P)$ profile, so for every wavelength and every source of absorption. Depending on the number of wavelength points and absorbing species, this can mean that the PINN has to calculate the solution in multiple batches, significantly reducing its speed. The advantage of our model lies within its versatility, as it is not trained on, e.g., specific molecules.

\subsection{Scattering PINN}

Since we cannot easily compare the spectra obtained from the scattering PINN to the ones from a traditional numerical method, which do not take scattering into account, evaluating its performance is harder to do.

\begin{figure}
   \centering
      \includegraphics[width=\columnwidth]{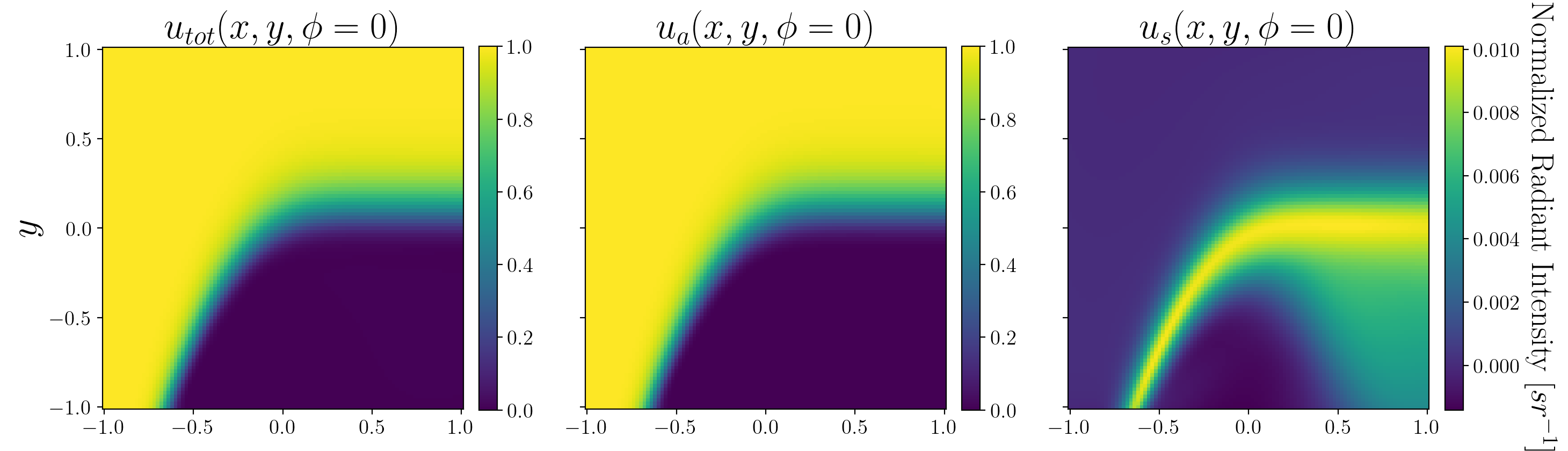}
      \vspace{0.00mm}
      \includegraphics[width=\columnwidth]{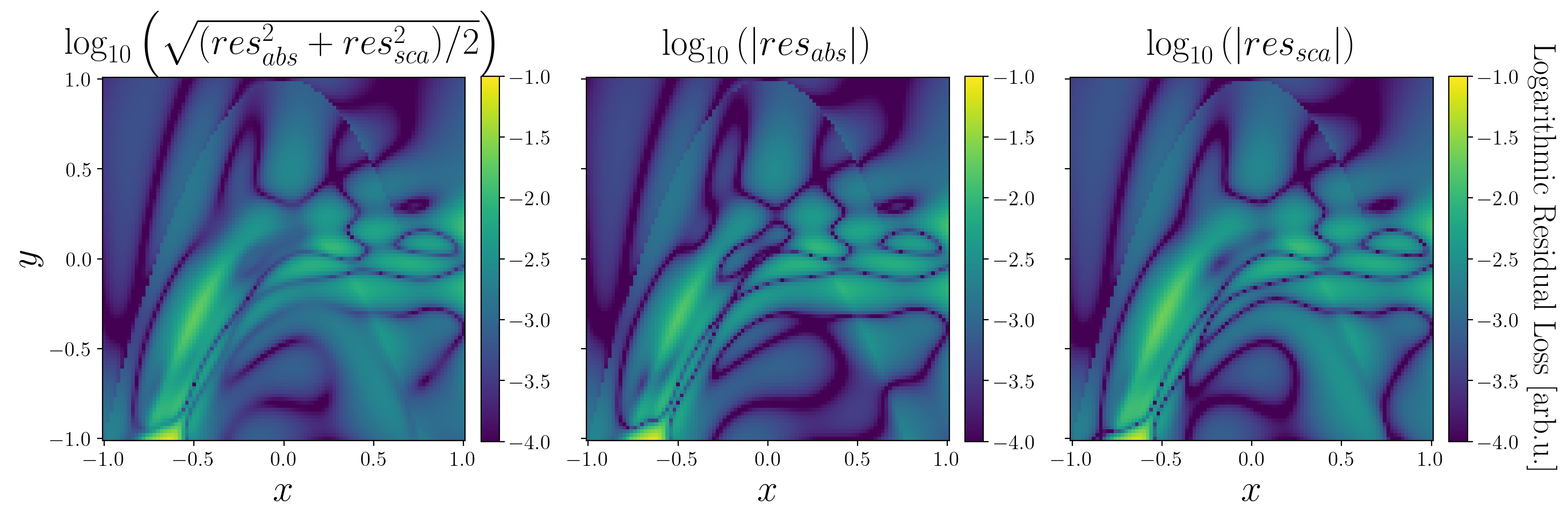}
   \caption{Example solution for the $\alpha(x,y)$ shown in Fig. \ref{fig:geometry} for the direction of the light in positive x-direction ($\phi=0$).\\
   Top: The complete solution (left), the solution if scattering would be treated as absorption (middle) and the difference between the two: the scattered light component (right), all in units of $u_0 := u(x=-1,\phi=0) = 1$.\\
   Bottom: The logarithm (log$_{10}$) of the respective residuals of the differential equations. The middle plot shows the residual of $u_a$ with respect to the RTE without scattering and the right one the residual of $(u_a + u_s)$ with respect to the complete RTE. The left plot is calculated from the two residuals by taking their root mean square.}
   \label{fig:scaPINN-sol+res}
\end{figure}

In Fig.\,\ref{fig:scaPINN-sol+res}, we can see an example solution with the two components, the absorbed $u_a$ and scattered light $u_s$ separated, as well as the corresponding residuals. These residuals show how well the RT equation is fulfilled and give us an estimate of the error. In the example, we can see residuals on the order of $10^{-2}$ spanning a significant fraction of the considered slab.

\begin{figure}
    \centering
    \includegraphics[width=\columnwidth]{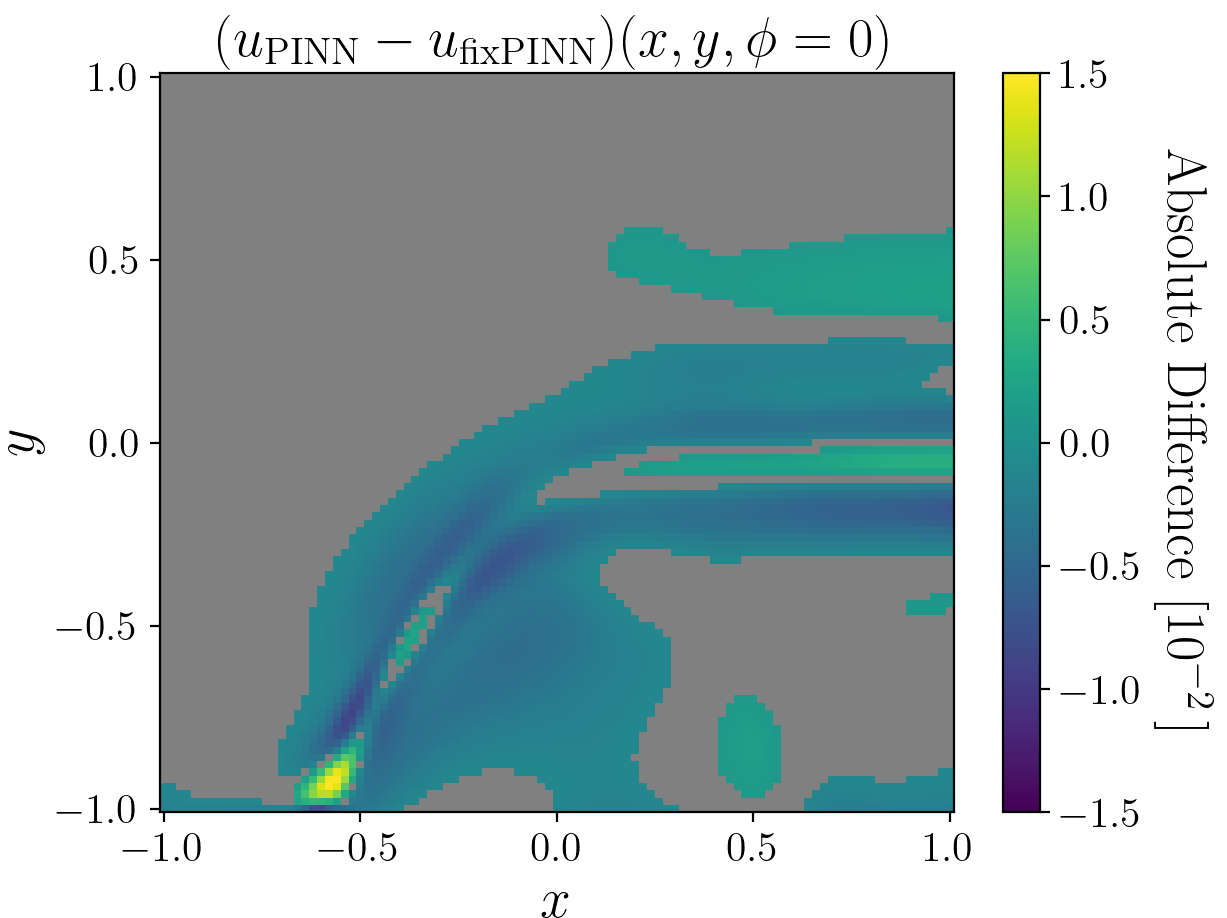}
    \caption{Comparison of the solution from the scattering PINN shown in Fig. \ref{fig:scaPINN-sol+res} with a higher-accuracy PINN with fixed parameters (apart from $x, y, \theta \; \mathrm{and} \; \Delta_*$). Grey indicates differences of less than 0.1\%.}
    \label{fig:pinnVSnum}
\end{figure}

Additionally, we compared our method to a fixed PINN for the same example, the result of which can be seen in Fig.\,\ref{fig:pinnVSnum}. By fixed PINN, we mean that all the parameters of this PINN have been fixed to certain values during training, apart from the spacial coordinates $x, y, \phi$ (and $\Delta_*$, which cannot be fixed because of the boundary condition at $y=-1$). Leaving the batch size the same should yield a solution of higher accuracy, as the fixed PINN has to explore a significantly smaller parameter space. Similarly to the absorption PINN, we can see errors of around 1\%.

\begin{figure}
    \centering
    \includegraphics[width=\columnwidth]{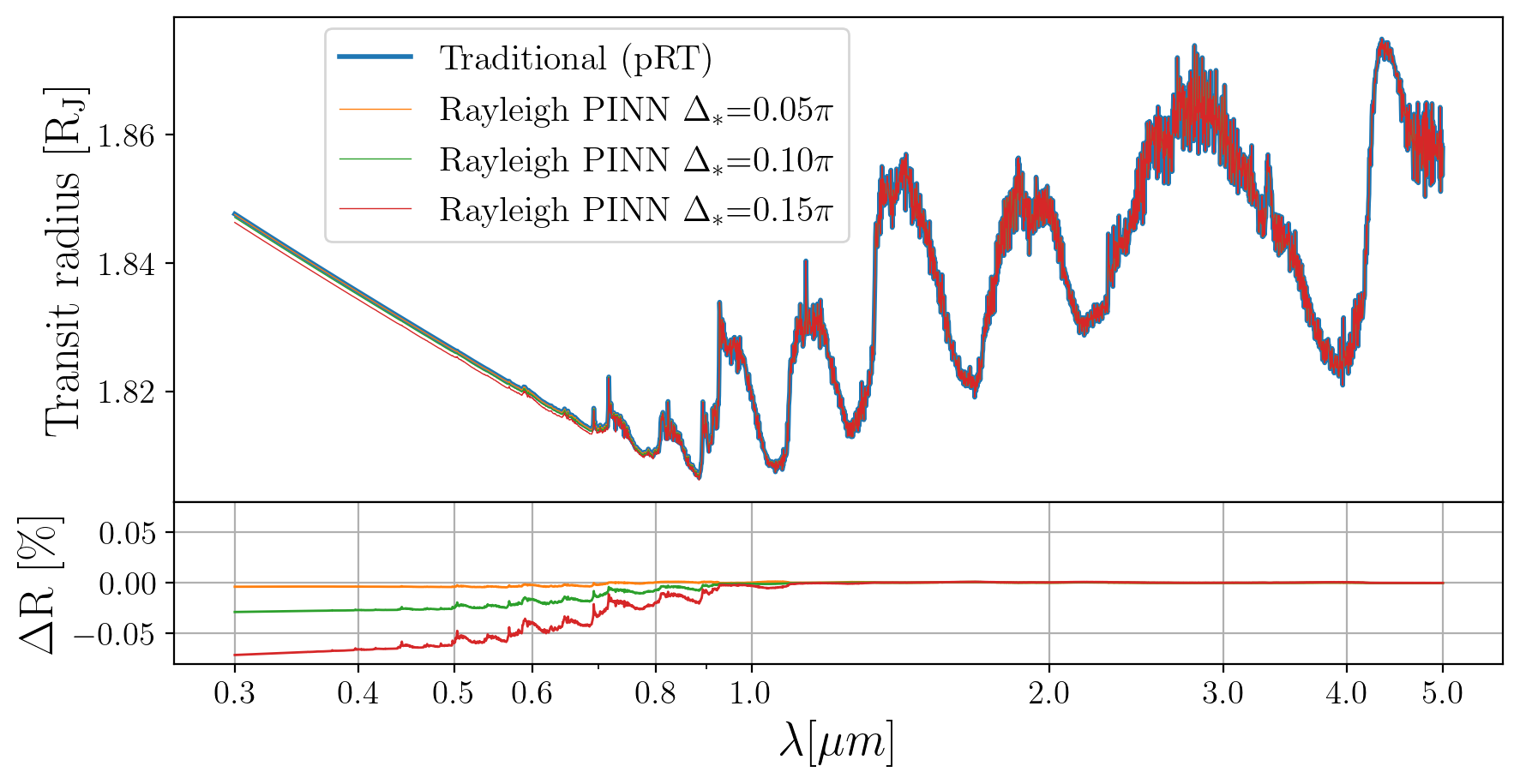}
    \caption{Example spectrum similar to Fig. \ref{fig:exampleSpec_absPINN} (without Na and K) with the solutions from petitRADTRANS \citep[][]{pRT} and the scattering PINN, each for different values of $\Delta_*$. We can see that even in the extreme case of $\Delta_* = 0.15\pi$ Rayleigh scattering barely influences the spectrum, when using our approximations.}
    \label{fig:exampleSpec_scaPINN}
\end{figure}

Finally, we want to explore the usefulness of this method by plotting an example spectrum (see Fig.\,\ref{fig:exampleSpec_scaPINN}). Even for the extreme case of $\Delta_* = 0.15\pi$, which would correspond to an exoplanet orbiting a sun-like star ($R_* = R_\odot$) at a distance of $a = 0.01$ AU, the Rayleigh scattering barely influences the spectrum. This is because our assumptions for calculating the scattering integral only hold for small angles. Thus, while the scattering PINN is able to find an accurate solution, the approximation of parameterizing our problem with only one angle is not sufficient to accurately model Rayleigh scattering for large angular extents of its host star, which is when it becomes relevant.

\section{Limitations} \label{sec:limitations}

The assumptions under which both the absorption and scattering PINN operate include isothermality of the atmosphere and a radius range of \( R_0 \in [0.2, 2.0] R_{\rm J} \).

In the case of the scattering PINN we further assume an orbital inclination of \( i = 90^\circ \). It also focuses narrowly on pressure-independent scattering cross-sections, characteristic of Rayleigh scattering from hydrogen and helium, and does not consider any absorbing species simultaneously being present in the atmosphere. Additionally, it lacks an exact lower boundary condition. The most significant shortcoming is the small angle approximation for the calculation of the scattering integral, which is ultimately insufficient to accurately model Rayleigh scattering.

Lastly, the fully-connected layers of both PINNs offer no interpretability, although the special architecture of the scattering PINN at least simplifies differentiating the diffuse light component from the total solution.

\section{Conclusions}

In this study, we have assessed the effectiveness of PINNs in simulating complex atmospheric phenomena like scattering, which are traditionally challenging and resource-intensive to model.

The absorption PINN demonstrated in this research exhibits a high level of precision in forecasting atmospheric transmission for transiting exoplanets, achieving an error margin within a few percent, even when confronted with unfamiliar non-differentiable $\alpha_a(P)$ profiles. This accuracy underscores the model's robustness and reliability in complex atmospheric simulations. However, it is worth noting that, in terms of computational speed, this method does not yet surpass the efficiency of existing numerical techniques. This opens up opportunities for further optimization and development in the realm of PINN methodologies to enhance their speed performance while maintaining their high accuracy. The use of other architectures than the simple MLP chosen in this work, could aid in achieving this. A 1D-CNN for the $\alpha_a$ inputs could, for example, significantly reduce the number of weights in the network.

The Rayleigh scattering PINN in this study, while showcasing potential, operates within certain constraints, as summarized in Section \ref{sec:limitations}, that currently limit its broader applicability. Having these caveats in mind, nevertheless, a parameterized PINN could prove effective in modeling scattering processes in the atmosphere with a minimal additional computational cost beyond the training phase.

In this work, we have laid out one possible path of how to present this complex problem to a neural network. Here we want to emphasize the key ingredients for successfully training a PINN:
\begin{itemize}
    \item especially when dealing with a complicated differential equation, one has to carefully consider all possible configurations of parameters, how they affect the absolute values of the respective residual and normalize all cases accordingly
    \item similar to the insight from \cite{Wang2022Causality}, it is important to respect causality when choosing weights for different parts of the loss function
    \item having a larger batch size generally improves the accuracy of a PINN
    \item incorporating physical knowledge into, e.g., the architecture of the PINN can significantly simplify finding the correct solution, especially if any non-differentiable boundaries are involved
\end{itemize}

However, further enhancements to this approach are necessary to calculate the correct solution to the RTE with scattering, where it becomes relevant for transit spectra. E.g., auxiliary PINNs \citep[][]{Yuan2022APINN, Riganti2023APINNforRTE} could aid in expanding the problem to a second angular dimension ($\theta$), while making it possible to evaluate the scattering integral without any numerical integration. Experimenting with alternative architectures like FNOs \citep[][]{Li2020FNO} could also make addressing these challenges more feasible.

In the future, the inclusion of specific pressure-temperature profiles can be achieved by simply extending the input parameters, though this might slightly impact the model's accuracy due to the increased dimensionality of the problem. Integrating other scattering sources, such as clouds with their variable $\alpha(P)$ profiles and complex scattering phase functions $\Phi(\theta, \theta')$, presents a more significant challenge. One major concern is the possible non-differentiability of cloud decks. But, as the absorption model's predictions for Na and K show, a PINN might be able to overcome such difficulties.

\section*{Acknowledgements}

We thank the referee for the helpful comments.
We also thank Lukas Heinrich, Nicole Hartman and the ORIGINS Data Science Lab (ODSL) for valuable discussions and hardware support during the development of the project, which was made possible through the funding of the ORIGINS Excellence Cluster by the Deutsche Forschungsgemeinschaft (DFG, German Research Foundation) under Germany’s Excellence Strategy - EXC 2094 - 390783311; in particular through ORIGINS Excellence Cluster's Seed Money Program.

This research has made use of the NASA Astrophysics Data System,
and the Python packages \textsc{NumPy}, \textsc{SciPy}, \textsc{Matplotlib}, and \textsc{PyTorch}.

\section*{Data Availability}

The code, based on \cite{Mishra2021}, is available at \url{https://github.com/DavidDahlbudding/AtmosphericScatteringPinn} (commit \href{https://github.com/DavidDahlbudding/AtmosphericScatteringPinn/tree/3c805f5f4a67be9caf76c644c91144fda956d6ea}{3c805f5}). The respective models and other data can be downloaded from \url{https://zenodo.org/doi/10.5281/zenodo.10727497}.



\bibliographystyle{mnras}
\bibliography{example} 




\appendix

\section{The Multi-Layer Perceptron}

A feedforward neural network or multilayer perceptron (MLP) is a versatile approximator that maps $\bm{x}$ to $\bm{y} = f(\bm{x},\bm{\theta})$, where $\bm{\theta}$ defines the free network parameters that are adjusted during training.

What makes these standard MLPs so versatile is just simple linear algebra: a linear mapping $\bm{W}$ (the weights), a displacement $\bm{b}$ (the biases) and, most importantly, a non-linear activation function $\sigma$. The output $\bm{y}$ of one such a layer $f_1$ can then be computed with:

\begin{equation}
    \bm{y} = f_1 (\bm{x}) = \sigma(\bm{W}\bm{x} + \bm{b}),
\end{equation}
where the activation function $\sigma$ is applied element-wise: $\sigma((x_1, x_2)^T) = (\sigma(x_1), \sigma(x_2))^T$. Multiple of these layers can be stacked on top of each other to be able to represent highly complex functions between the input $\bm{x}$ and the output $\bm{y} = f_{n} \circ \cdots \circ f_2 \circ f_1 (\bm{x})$ (where $\circ$ denotes a function composition: $f_2 \circ f_1 (\bm{x}) = f_2(f_1 (\bm{x}))$) \citep[e.g.][]{Goodfellow2016}. The layers between the input and the output layer are called the hidden layers and the number of hidden layers is called the depth of the network, while the number of nodes per layer is called the width. This simple architecture is also mostly used in this paper.

To train a neural network, we need another important function, the so-called objective or loss function. This function should be minimized to achieve a desired goal (e.g., the mean squared error when fitting a function to given data points). Then, the gradient of the loss function with respect to the weights and biases can be cumulatively calculated for a batch of data using automatic differentiation, and the neural network can "move" towards a minimum by adjusting its parameters in small steps in the negative direction of this gradient.

\section{The Scattering Integral} \label{app:sca_int}

As mentioned in Section \ref{sec:setup} we need to make a few approximations valid for small angular extents of the star for the problem to be parameterizable by only one angle ($\phi$). Here, we choose different symmetries for the two components of the radiant flux: $u_a$ and $u_s$. Because they are added together to get the total solution to the RTE, we can calculate the total integral by adding the two integrals for each component:

\begin{equation}
    \int \Phi \, u \, \dd\Omega = \int \Phi \, u_a \, \dd\Omega + \int \Phi \, u_s \, \dd\Omega
\end{equation}

For the absorption component $u_a$ we choose the conventional spherical (local) coordinate system, such that the angle $\phi \in [-\pi, \pi]$ lies in the x-y-plane and $\theta \in [0,\pi]$ is the angle relative to the z-axis. If we now take a fixed $\phi$, let $\theta$ vary and expand this into a plain, the intersection of this plain with the sphere of an atmospheric layer is a circle. For small angular extents of the star each light ray for varying $\theta$ travels roughly the same path through this circle, which is why we assume the symmetry of $u_a(\phi, \theta) \approx u_a(\phi, \theta=\frac{\pi}{2})$.

We can parameterize $\theta'$ in terms of $\phi'$ such that we can integrate over the complete solid angle of the star with angular extent $2 \Delta_*$, for which $u_a \neq 0$, by using the law of cosines with $\theta = \frac{\pi}{2}$:

\begin{equation}
    \mathrm{cos}\Delta = \mathrm{sin}\theta' \mathrm{cos}(\phi-\phi').
\end{equation}

Inserting this into the scattering phase function $\Phi(\Delta)$ (\ref{eq:scatteringPhaseFunction}) and the integration limits, yields:

\begin{equation}
\begin{split}
    &\int_{-\Delta_*}^{\Delta_*} \dd\phi' \, u_a(\phi') \int_{-\theta'(\phi')}^{\theta'(\phi')} \dd\theta' \, |\mathrm{sin}\theta'| \, \Phi(\Delta) =\\
    &\int_{-\Delta_*}^{\Delta_*} u_a(\phi') \, \frac{3}{2} \sqrt{1-a^2} \left( \mathrm{cos}^2(\phi-\phi') \left( \frac{1-a^2}{3} - 1\right) - 1 \right) \dd\phi',\\
    & \mathrm{where} \; a := \frac{\mathrm{cos}\Delta_*}{\mathrm{cos}\phi'}
\end{split}
\end{equation}

For exploiting a symmetry of the scattering component we have to rotate our local coordinate system. The $\phi_s$, coordinate now lies in the y-z-plane and $\theta_s$ is the angle relative to the x-axis. The absorption component $u_a$ therefore is a concentrated beam of light around $\theta_s = 0$ for small $\Delta_*$. Since the scattering phase function is only dependent on the angular distance $\Delta$, relative to $\theta=0$, we get $\Delta = \theta$. Hence we can approximate, while allowing for some slight asymmetry: $u_s(\theta_s, \phi_s \lessgtr 0) \approx u_s(\theta_s, \phi=\pm\frac{\pi}{2})$. The integral can therefore be calculated using:

\begin{equation}
\begin{split}
    \iint \Phi(\Delta) \, u_s(\theta_s'&, \phi_s') \, |\mathrm{sin}\theta_s'| \, \dd\theta_s' \, \dd\phi_s' =\\
    \int_{0}^{\pi} |\mathrm{sin}\theta_s'| \, \dd\theta_s' &\left[ u_s(\theta_s', \phi_s'=-\frac{\pi}{2}) \int_{-\pi}^0 \Phi(\Delta) d\phi_s' + \right.\\
    &\left. \; u_s(\theta_s', \phi_s'=\frac{\pi}{2}) \int_{0}^{\pi} \Phi(\Delta) d\phi_s' \right]\,.
\end{split}
\end{equation}

The integral over $\phi_s'$ can again be integrated analytically by inserting the law of cosines (assuming $\phi_s = \pm \frac{\pi}{2}$) into the scattering phase function $\Phi(\Delta)$ (\ref{eq:scatteringPhaseFunction}):

\begin{equation}
\begin{split}
    \int_0^{\pm\pi} & \Phi(\Delta) d\phi_s' =\\
    \pm \frac{3}{4} & \left[ \pi \left( 1 + \mathrm{cos}\theta_s \mathrm{cos}\theta_s' \pm \frac{1}{2} \mathrm{sin}\theta_s \mathrm{sin}\theta_s'\right) \right.\\
    & \left. \mp \mathrm{sgn}(\phi_s) \mathrm{sin}(2\theta_s) \mathrm{sin}(2\theta_s') \right]\,.
\end{split}
\end{equation}

The numerical integration employs the Trapezoidal rule with a total of 19 sample points in the interval $[-\pi,\pi]$, where 5 of these points are equally spaced in the interval $[-\Delta_*, \Delta_*]$ and the remaining intervals $[-\pi, -\Delta_*]$ and $[\Delta_*, \pi]$ each have another 7 equidistant sample points.

\section{Residual Loss for Scattering PINN} \label{app:res_loss}

As mentioned in Section \ref{sec:dataGen}, we have to divide the gradients in the RTE (Eq.\,\ref{eq:rte}) with the respective scale lengths $l_x$ and $l_y$. Given in cm, these parameters are quite large and can therefore diminish the value of the residual loss, especially for low values of $\alpha$. Hence, we multiply the whole equation with $l_x$:

\begin{equation}
\begin{split}
    \mathrm{cos}(\phi) \frac{\dd u}{\dd x} &+ \mathrm{sin}(\phi) \frac{l_x}{l_y} \frac{\dd u}{\dd y} + l_x \alpha_s \cdot u\\
    - \frac{l_x \alpha_s}{4\pi} &\cdot \int \Phi(\theta, \phi, \theta', \phi') u(\theta', \phi') d\Omega' = \mathrm{res}_1.
    \label{eq:scaRes1}
\end{split}
\end{equation}

The remaining problem is that the value of $\frac{l_x}{l_y}$ can be on the order of 10, possibly inflating the value of the residual loss for a certain set of parameters. Because of this, we further normalize this formulation of the residual with $\mathrm{norm} = |\mathrm{cos}(\phi)| + \frac{l_x}{l_y} |\mathrm{sin}(\phi)|$.

The second formulation is similar to the one for the absorption PINN, where we divide the RTE by $\alpha_s$:

\begin{equation}
\begin{split}
    \frac{\mathrm{cos}(\phi)}{l_x \alpha_s} \frac{\dd u}{\dd x} &+ \frac{\mathrm{sin}(\phi)}{l_y \alpha_s} \frac{\dd u}{\dd y} + u\\
    - \frac{1}{4\pi} &\cdot \int \Phi(\theta, \phi, \theta', \phi') u(\theta', \phi') d\Omega' = \mathrm{res}_2.
    \label{eq:scaRes2}
\end{split}
\end{equation}

The complete residual is then again the minimum of the two formulations:

\begin{equation}
    \mathrm{res} = \mathrm{min} \left( \frac{\mathrm{res}_1}{\mathrm{norm}}, \mathrm{res}_2 \right).
\end{equation}


\bsp	
\label{lastpage}
\end{document}